%% file: VPpaper.tex
\documentclass[final,3p,twocolumn,longtitle]{elsarticle}

\usepackage{graphicx}

\usepackage{amssymb}

\journal{Physics Letters B}

\newcommand{\GeV}{\,\mathrm{GeV}}
\newcommand{\GeVc}{\GeV/c}
\newcommand{\GeVcc}{\GeV/c^2}
\newcommand{\MeV}{\,\mathrm{MeV}}

\newcommand{\MeVcc}{\MeV/c^2}

\begin{document}

\begin{frontmatter}



\title{Measurement of cross sections of exclusive $e^+ e^- \rightarrow VP$
processes \\ at $\sqrt{s}=10.58$ GeV \\ \vskip8pt
{\small (Belle Collaboration)}}

\renewcommand\theaffn{\arabic{affn}}

\include{author}


\date{\today}


\begin{abstract}
The cross sections for the reactions 
$e^+e^- \rightarrow \phi\eta, \quad \phi\eta', \quad \rho\eta, \quad \rho\eta'$
have been measured using a data sample of 516 fb$^{-1}$ collected with the
Belle detector at the KEKB asymmetric-energy $e^+e^-$ collider. The
corresponding values of the cross sections are:
$1.4 \pm 0.4 \pm 0.1$ fb $(\phi\eta)$,
$5.3 \pm 1.1 \pm 0.4$ fb $(\phi\eta')$,
$3.1 \pm 0.5 \pm 0.1$ fb $(\rho\eta)$ and
$3.3 \pm 0.6 \pm 0.2$ fb $(\rho\eta')$.
The energy dependence of the cross sections is presented using Belle
measurements together with those of CLEO and BaBar.
\end{abstract}


\end{frontmatter}



High-statistics data samples accumulated at B-factories allow a study of rare
exclusive two-body processes in $e^+e^-$ annihilation. An example of such a
highly suppressed process is the reaction $e^+e^- \rightarrow VP$, where $V$
and $P$ stand for Vector and Pseudoscalar mesons, respectively. It has been
observed~\cite{DCHARM1, DCHARM2} that double charm production in
$e^+e^- \rightarrow J/\psi \eta_c $ has an unexpectedly high cross section.
The basic diagram for double charm production is very similar to the one
describing $e^+e^- \rightarrow \phi \eta(\eta')$ where c quarks are replaced
by s quarks. Thus, comparison of the two reactions may contribute to better
understanding of the underlying physics. In addition, we investigate the
processes $e^+e^- \rightarrow \rho \eta (\eta')$, which also belong to the
$VP$ class with a different isospin configuration and light quarks only.

Some of the $e^+e^- \rightarrow VP$ reactions have previously been measured at
different center-of-mass (CM) energies: in the DM1 experiment at $\sqrt{s}$
between $1.4$ and $2.18\GeV$~\cite{DM1}, in the CLEO experiment at
$\sqrt{s} = 3.67\GeV$~\cite{CLEO1} and by the BaBar collaboration at
$\sqrt{s}$ between $1$ and $3\GeV$ using initial-state radiation
(ISR)~\cite{BaBar0} and at $\sqrt{s} = 10.58\GeV$~\cite{BaBar1}. The cross
section of the process $e^+e^- \rightarrow \rho \eta$ was also measured by the
BES collaboration at $\sqrt{s} = 3.65\GeV$~\cite{BES}. The reaction
$e^+e^- \rightarrow \phi \eta'$ has not yet been observed, nevertheless the
upper limit on its cross section set by CLEO~\cite{CLEO1} can be useful for
discrimination between models that predict different energy dependences.

The QCD-based models predict the energy dependence for the process
$e^+e^- \rightarrow VP$ to be $1/s^4$~\cite{QCD1, QCD2} while the cross section
for the process $e^+e^- \rightarrow \phi \eta$ measured by CLEO and BaBar
favors a $1/s^3$ dependence. The form factor for the process
$e^+e^- \rightarrow VP$ is expected to have a $1/s$ dependence~\cite{LOPEZ}.
Recently theoretical calculations of $e^+e^- \rightarrow VP$ cross sections
have been published, which use the light cone approach~\cite{LC, Likhoded}.
Predictions are given for two values of $\sqrt{s}$: $3.67$ and $10.58\GeV$.
The authors of Ref.~\cite{LC} claim that their results favor a $1/s^3$
dependence. In Ref.~\cite{Likhoded} $\sigma \sim 1/s^4$ is expected in the
limit $s \rightarrow \infty$.

%
The analysis presented here is based on data taken at the $\Upsilon(4S)$
($\sqrt{s} = 10.58\GeV$) with the Belle detector at the KEKB asymmetric-energy
$e^+e^-$ collider~\cite{KEKB}. The total integrated luminosity of the
on-resonance sample used in the analysis is 516 fb$^{-1}$. To check that the
processes $e^+e^- \rightarrow VP$ are due to a single-photon annihilation and
that the hadronic decay of the $\Upsilon(4S)$ does not give a significant
contribution, we use a 58 fb$^{-1}$ data sample collected 60 MeV below the
resonance peak. All the observed off-resonance signals are consistent with
those at the $\Upsilon(4S)$ resonance within statistical errors. We use these
data to set upper limits for the branching ratios of
$\Upsilon(4S) \rightarrow VP$.

A detailed description of the Belle detector is given elsewhere~\cite{BELLE}.
We mention here only the detector components essential for the present
analysis. Charged tracks are reconstructed from hit information in the central
drift chamber (CDC) located in a 1.5 T solenoidal magnetic field. Trajectories
of charged particles near the collision point are provided by a silicon vertex
detector (SVD). Photon detection and energy measurements are performed with a
CsI(Tl) electromagnetic calorimeter (ECL). Identification of charged particles
is based on the information from the time-of-flight counters (TOF) and silica
aerogel Cherenkov counters (ACC). The ACC provides good separation between
kaons and pions or muons at momenta above $1.2\GeVc$. The TOF system consists
of 128 plastic scintillation counters and is effective in $K/\pi$ separation
for tracks with momenta below $1.2\GeVc$. Low energy kaons are also identified
using specific ionization  $(dE/dx)$ measurements in the CDC.

In order to identify hadrons, for each of the three hadron types
$i\ (i = \pi,\ K\ \mathrm{and}\ p)$ a likelihood $L_i$ is formed using
information from the ACC, TOF, and $dE/dx$ measurements from the CDC. Kaons are
selected with the requirement $L_K/(L_K + L_\pi) > 0.6$, which has an efficiency
of 90\% and 6\% probability to misidentify a pion as a kaon. All charged tracks
that are not identified as kaons are considered to be pions.

Signal candidates are selected in two steps. Initially, events with low
multiplicity are selected by requiring that the number of charged tracks in an
event be two or four with zero net charge and each track have a momentum
transverse to the beam axis ($p_t$) larger than $0.1\GeVc$; and that each
track extrapolate to the interaction point (IP) within 1 cm transversely and
within 5 cm along the beam direction. To suppress background from Bhabha and
$\mu^+\mu^-$ events, the sum of the absolute values of momenta of the first
and second highest momentum tracks are required to be less than $9\GeVc$.
At least one track with $p_t$ above $0.5\GeVc$ is required. Beam-related
background is suppressed by requiring that the position of the reconstructed
event vertex be less than 0.5 cm from the IP in the transverse direction and
less than 3 cm from the IP along the beam direction.

Photons are defined as ECL clusters with energy deposits above $200\MeV$ that
are not associated with charged tracks. Neutral pion candidates are formed from
pairs of photons with invariant masses in the range 120 to $150\MeVcc$. The
two-photon invariant mass resolution in the mass region of the $\pi^0$ is about
$6\MeVcc$. For the $\eta \rightarrow \gamma \gamma$ reconstruction we use only
photons that do not form a $\pi^0$ candidate with any other photon. The
invariant mass of $\eta \rightarrow \gamma \gamma$ candidates should lie in
the range $0.5-0.6\GeVcc$. The two-photon invariant mass resolution in the mass
region of the $\eta$ meson is about $20\MeVcc$.
 
After preselection we apply the following requirements to extract the exclusive
$VP$ final states:

\begin{itemize}
\item The difference between the energy of $VP$ candidates in the CM frame and
$\sqrt{s}$ of KEKB should be between $-0.3$ and $+0.2\GeV$;\footnote{
This cut allows for low energy photon radiation up to $0.3\GeV$ to be present.}
\item The angle between $V$ and $P$ candidates in the CM frame should be larger
than 175 degrees.
\end{itemize}

\begin{figure}
\vspace*{\baselineskip}
{\hspace*{0.12\columnwidth}\bf a) \hspace*{0.42\columnwidth} b) \hfill ~} \vspace*{-2\baselineskip} \\
\includegraphics[width=0.48\columnwidth]{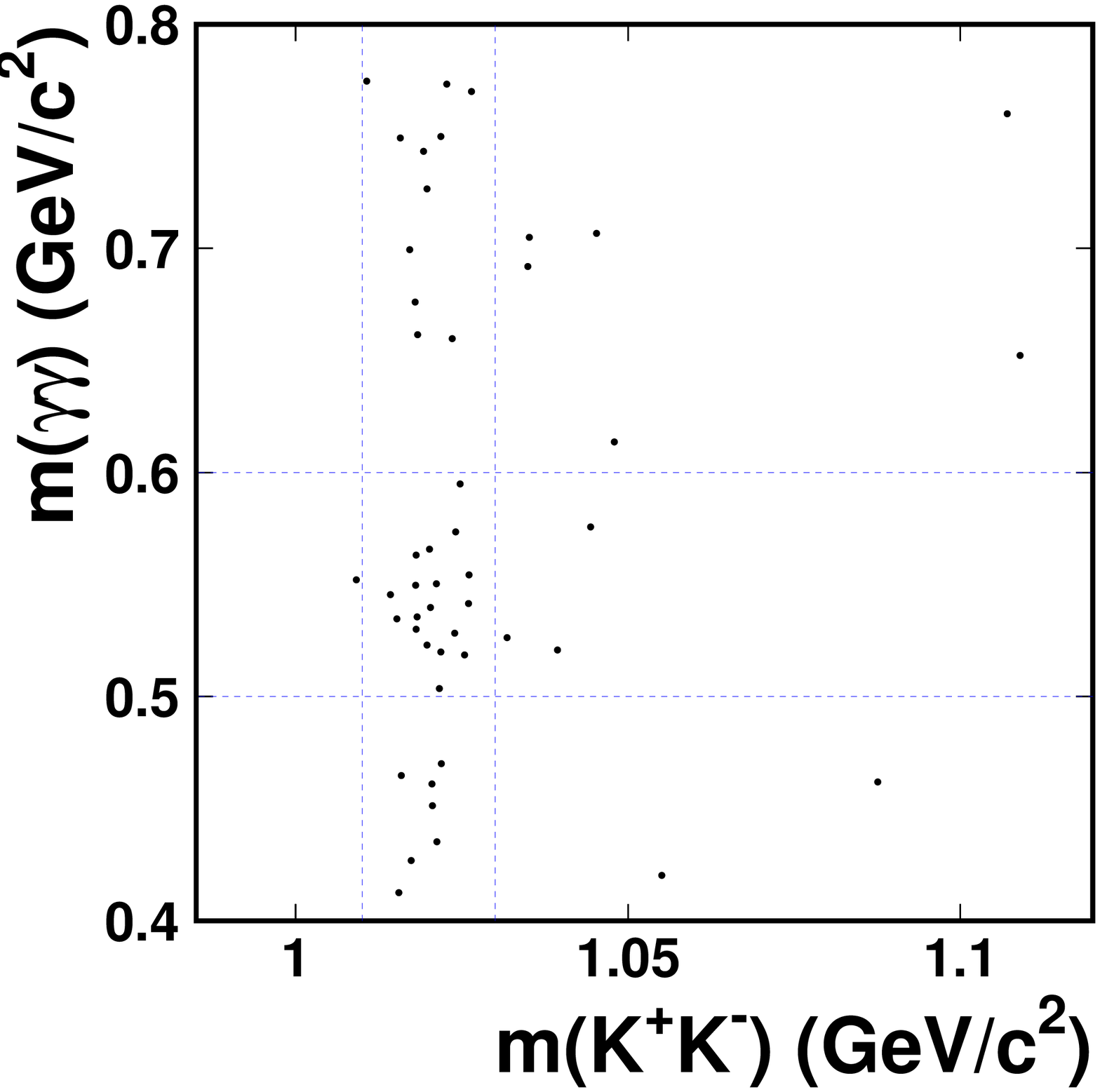}
\includegraphics[width=0.48\columnwidth]{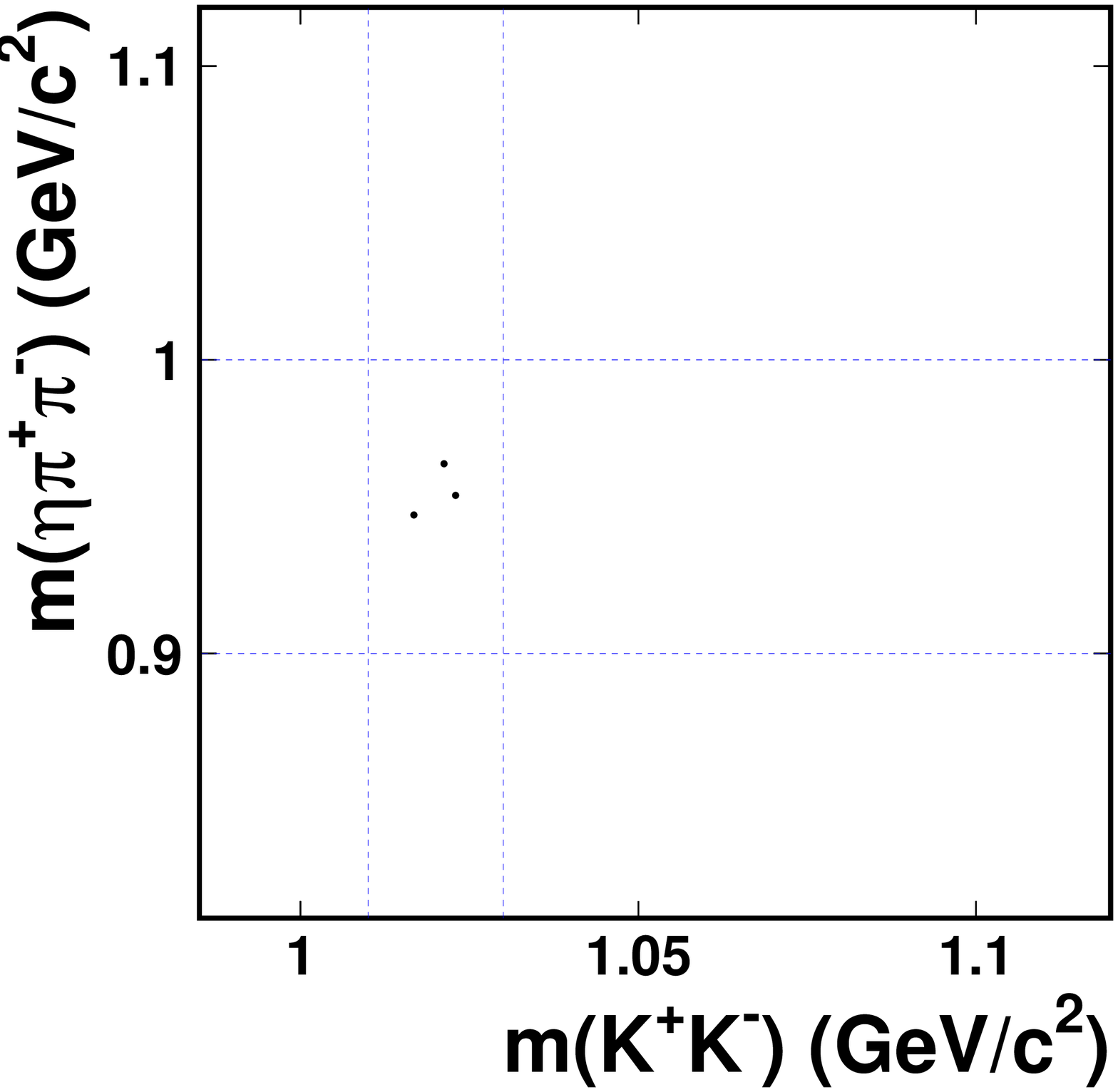} \vspace*{\baselineskip} \\
{\hspace*{0.12\columnwidth}\bf c) \hspace*{0.42\columnwidth} d) \hfill ~} \vspace*{-2\baselineskip} \\
\includegraphics[width=0.48\columnwidth]{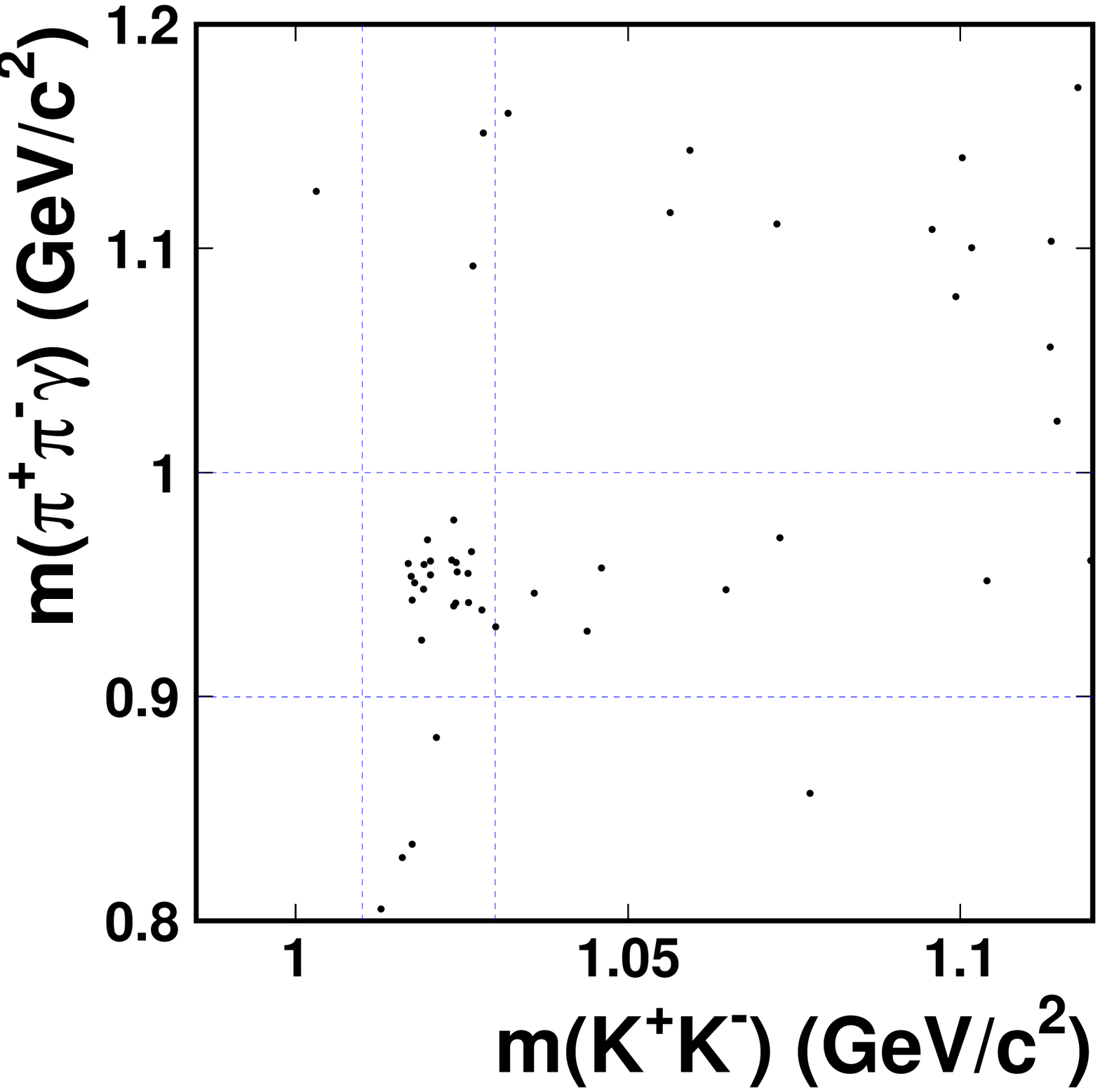}
\includegraphics[width=0.48\columnwidth]{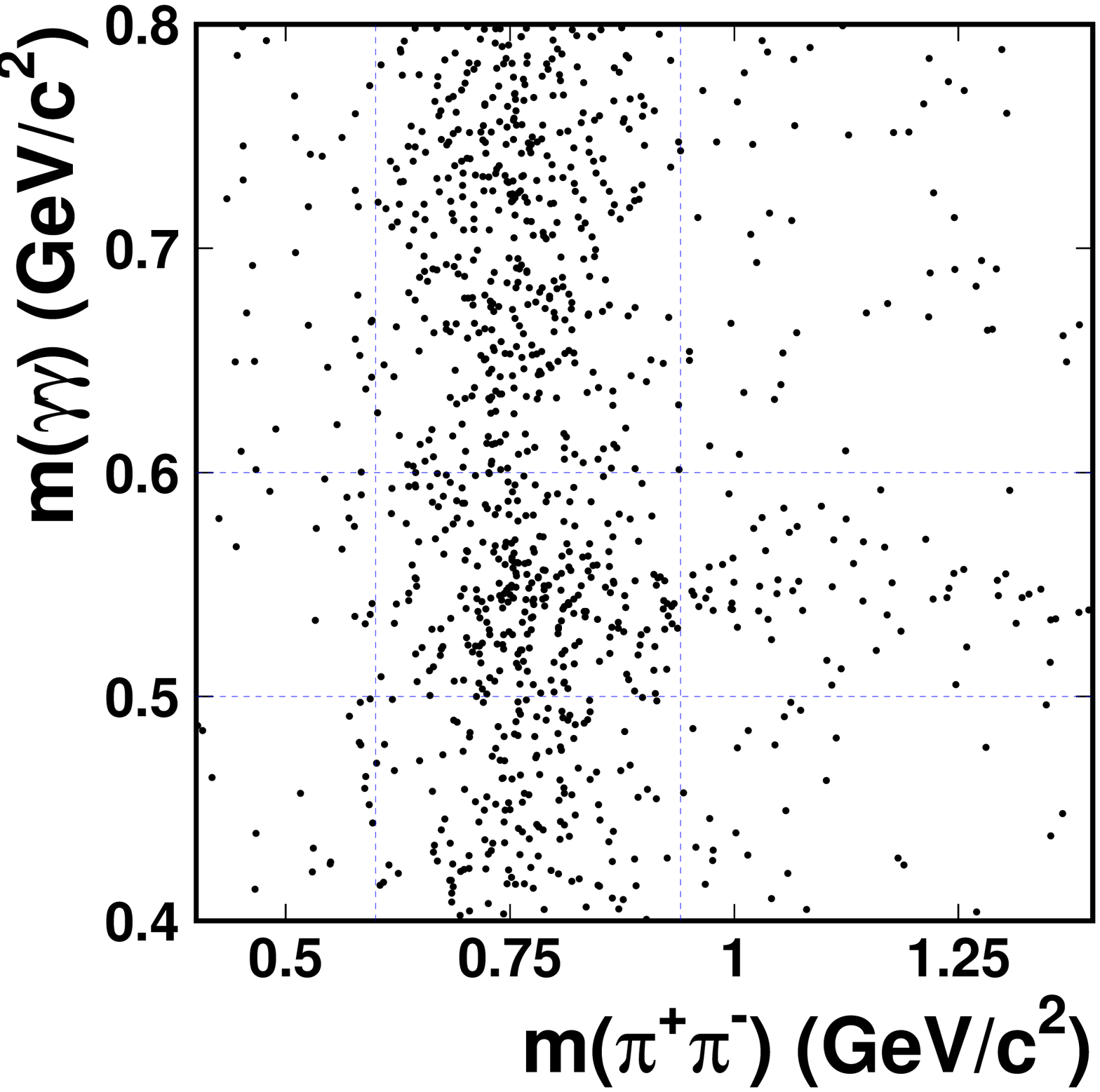} \vspace*{\baselineskip} \\
{\hspace*{0.12\columnwidth}\bf e) \hspace*{0.42\columnwidth} f) \hfill ~} \vspace*{-2\baselineskip} \\
\includegraphics[width=0.48\columnwidth]{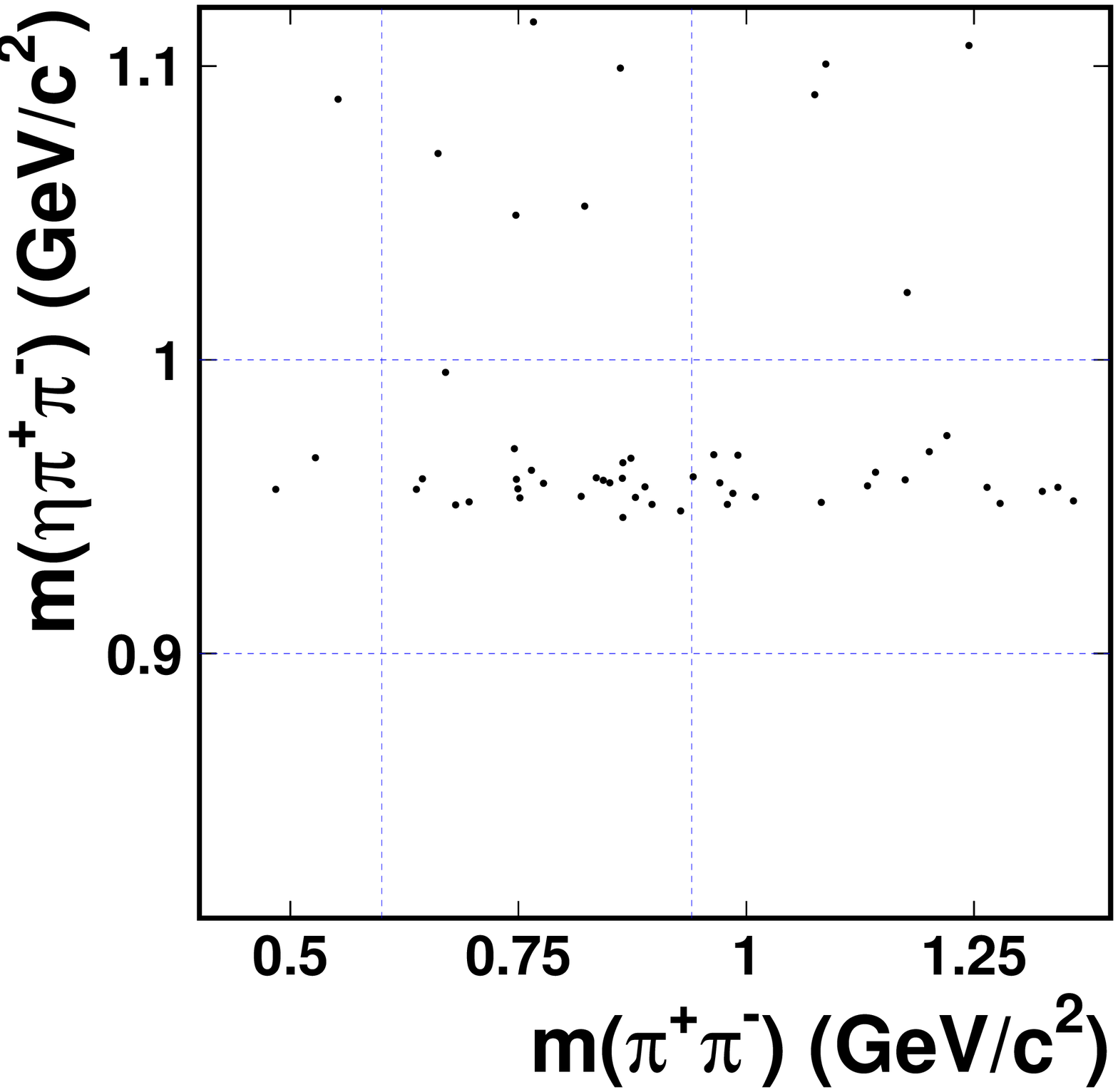}
\includegraphics[width=0.48\columnwidth]{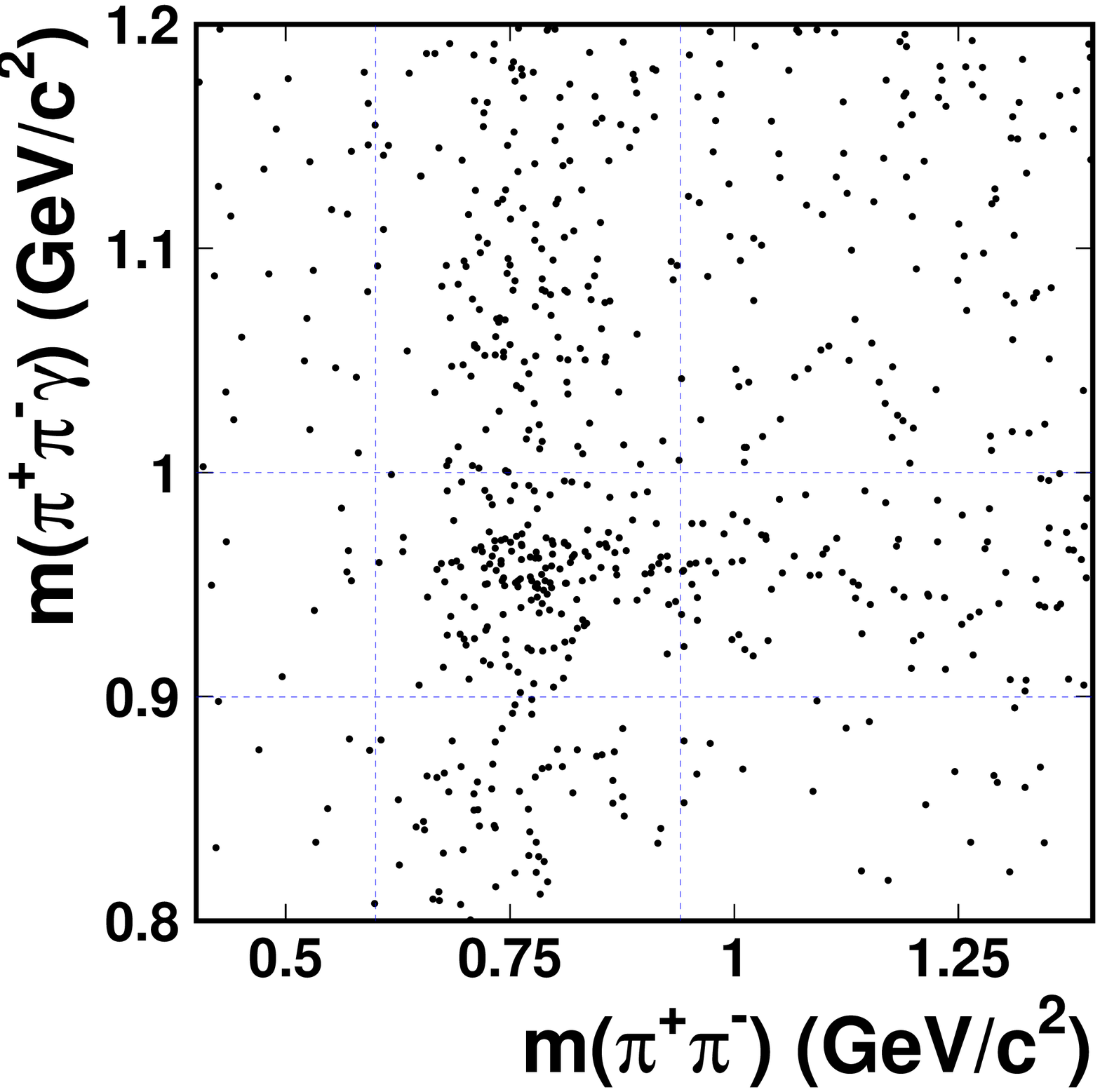}
\caption{\label{fig:scatter}Scatter plots for the processes 
{\bf a)} $e^+e^- \rightarrow \phi\eta \rightarrow K^+K^-\gamma\gamma$,
{\bf b)} $e^+e^- \rightarrow \phi\eta' \rightarrow K^+K^- \eta\pi^+\pi^-$,
{\bf c)} $e^+e^- \rightarrow \phi\eta' \rightarrow K^+K^- \pi^+\pi^-\gamma$, 
{\bf d)} $e^+e^- \rightarrow \rho\eta \rightarrow \pi^+\pi^- \gamma\gamma$,
{\bf e)} $e^+e^- \rightarrow \rho\eta' \rightarrow \pi^+\pi^- \eta \pi^+\pi^-$,
{\bf f)} $e^+e^- \rightarrow \rho\eta' \rightarrow \pi^+\pi^- \pi^+\pi^-\gamma$.
The dotted lines show the mass ranges used for one-dimensional projections.}
\end{figure}

We consider the following decay modes of vector and pseudoscalar mesons:
$\phi(1020) \rightarrow K^+ K^-$, $\rho \rightarrow \pi^+ \pi^- $, 
$\eta \rightarrow \gamma \gamma$, $\eta' \rightarrow \pi^+\pi^-\gamma$, 
$\eta' \rightarrow \eta \pi^+ \pi^-$. After the application of the requirements
listed above we observe significant concentrations of events in the scatter
plots near the masses of corresponding vector and pseudoscalar mesons. These
scatter plots are shown in Figs.~\ref{fig:scatter}(a-f).

A two-dimensional unbinned likelihood fit is applied to extract the signal
yields for the above reactions. We assume the mass distributions of vector and
pseudoscalar particles to be uncorrelated; thus the distributions in the
scatter plots of Fig.~1 can be represented as the product of two
one-dimensional probability density functions (PDF), one for each
dimension.\footnote{Two-dimentional function thus could be written as: \\
$f(m_1,m_2)=( A\cdot s1 + B\cdot b1 )\cdot ( C\cdot s2 + D\cdot b2 )$,
where $s1(m_1)$ and $b1(m_1)$ are one-dimentional signal and background
functions for vector particle, $s2(m_2)$ and $b2(m_2)$ -- for pseudoscalar
particle, $A$, $B$, $C$ and $D$ are free fitting parameters.}
To fit the $\phi(1020) \rightarrow K^+K^-$ and $\rho \rightarrow \pi^+\pi^-$
invariant mass distributions we use a non-relativistic Breit-Wigner function.
The $\eta \rightarrow \gamma\gamma$ and
$\eta' \rightarrow \pi^+\pi^-\gamma, \quad \eta\pi^+\pi^-$ invariant mass
distributions are fitted with Gaussians. The background for the $K^+K^-$
system is described by the product of the threshold function
$(m(K^+K^-)-m_0)^\alpha$ and a first-order polynomial, where $\alpha$ is a free
parameter and $m_0=2m_{K^+}$. The fit range for $m(K^+K^-)$ extends from $m_0$
to $1.12\GeVcc$. The backgrounds for the $\pi^+\pi^-$, $\gamma \gamma$,
$\eta \pi^+ \pi^-$ and $\pi^+\pi^-\gamma$ systems are parameterized with
first-order polynomials. The fit ranges are 0.4--1.4$\GeVcc$, 0.4--0.8$\GeVcc$,
0.83--1.12$\GeVcc$ and 0.8--1.2$\GeVcc$ for $\pi^+\pi^-$, $\gamma \gamma$,
$\eta \pi^+ \pi^-$ and $\pi^+\pi^-\gamma$, respectively. The two-dimensional
fitting functions for the scatter plots are the sum of products of the
corresponding one-dimensional signal and background functions. The mean values
(masses) of the signal functions are fixed at PDG values~\cite{PDG} while their
widths are fixed to the values obtained from the corresponding inclusive
spectra in the data. The significance of the fit is defined as
$\sqrt{-2\mathrm{ln}(L_0/L_\mathrm{max})}$, where $L_0$ and $L_\mathrm{max}$
are the likelihood values returned by the fit with signal yield fixed to zero
and at its best fit value. The signal yields obtained from this fit procedure
and the significance of the fits for all processes are presented in
Table~\ref{tab:result}.

\begin{figure}
\vspace*{\baselineskip}
{\hspace*{0.12\columnwidth}\bf a) \hspace*{0.42\columnwidth} b) \hfill ~} \vspace*{-2\baselineskip} \\
\includegraphics[width=0.48\columnwidth]{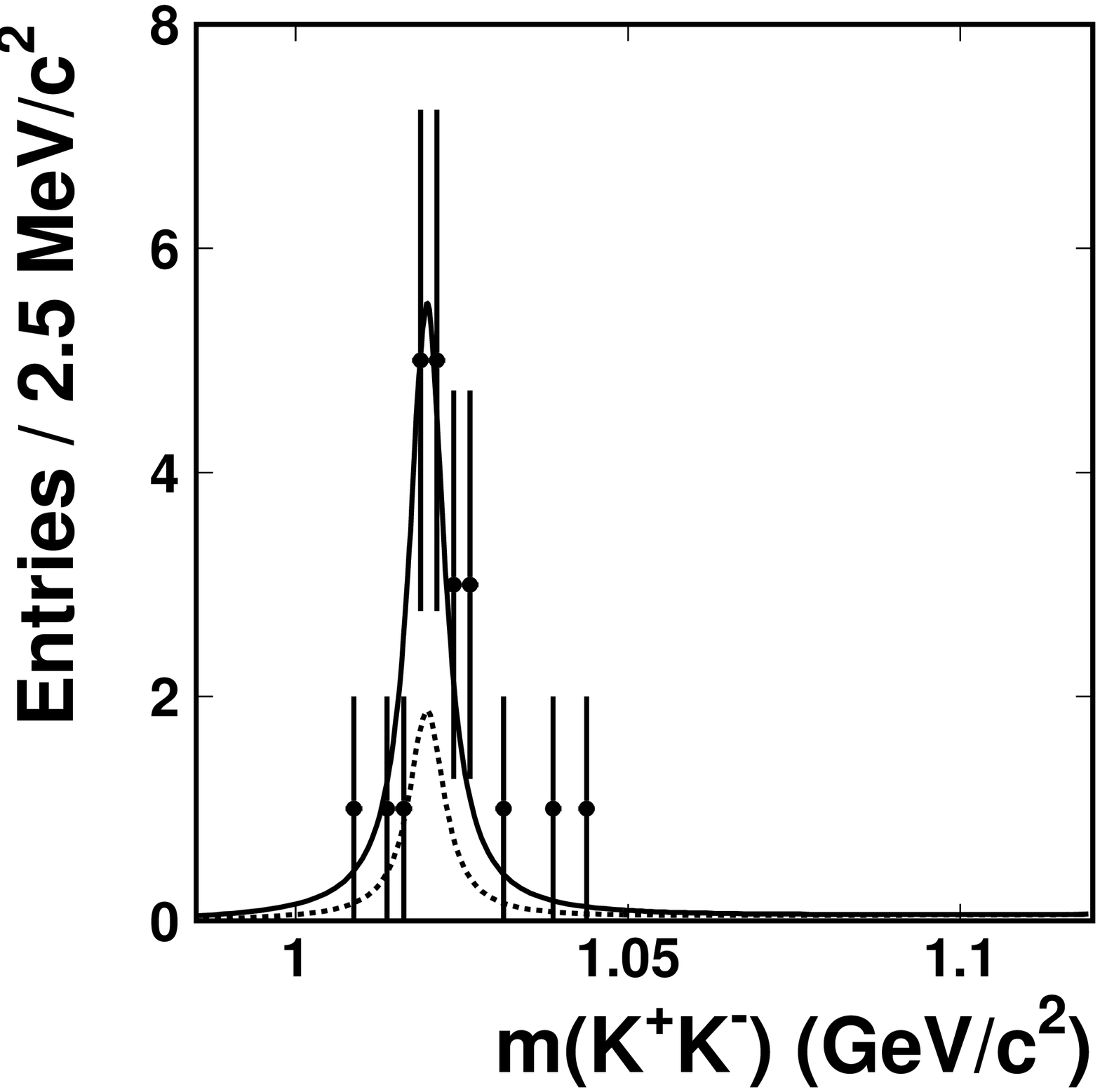}
\includegraphics[width=0.48\columnwidth]{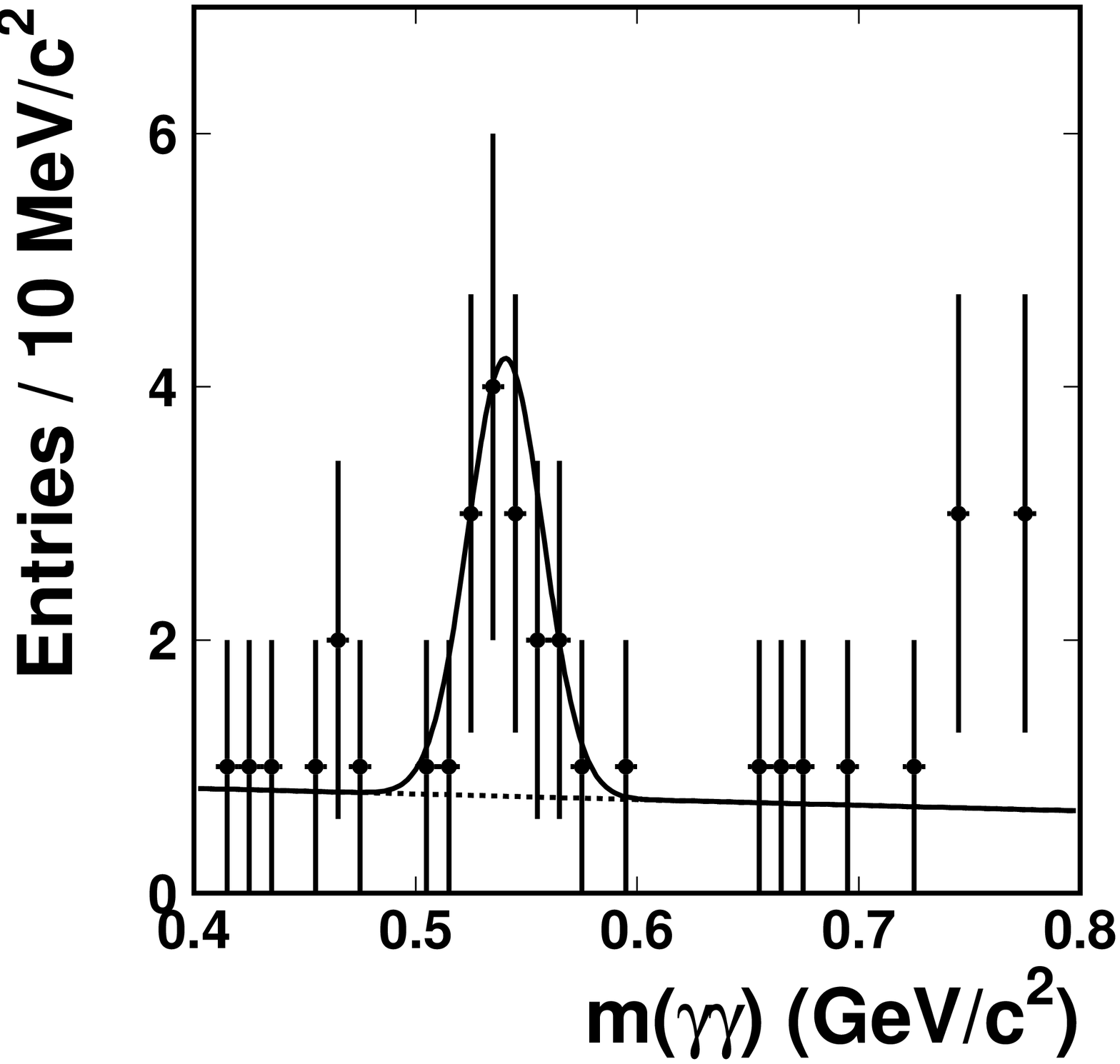}
\caption{The mass projections for the scatter plot
$m(K^+K^-)$ vs. $m(\gamma\gamma)$ onto
{\bf a)} $K^+K^-$ and
{\bf b)} $\gamma\gamma$
for the reaction $e^+e^- \rightarrow \phi\eta \rightarrow K^+K^-\gamma\gamma$.
The solid curves show the result of the two-dimensional fit,
the dotted curves show the background contamination.}
\end{figure}

\begin{figure}
\vspace*{\baselineskip}
{\hspace*{0.12\columnwidth}\bf a) \hspace*{0.42\columnwidth} b) \hfill ~} \vspace*{-2\baselineskip} \\
\includegraphics[width=0.48\columnwidth]{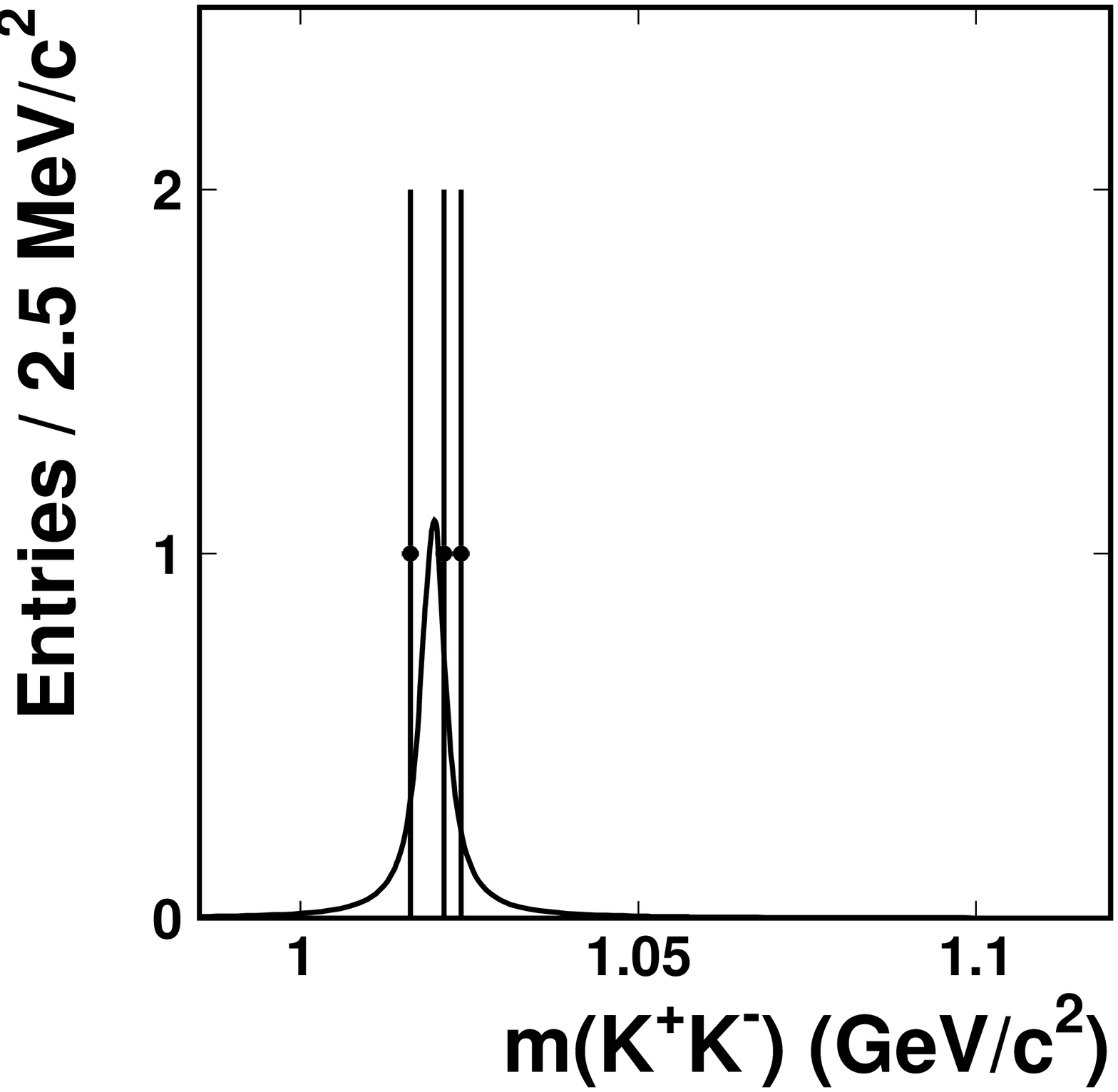}
\includegraphics[width=0.48\columnwidth]{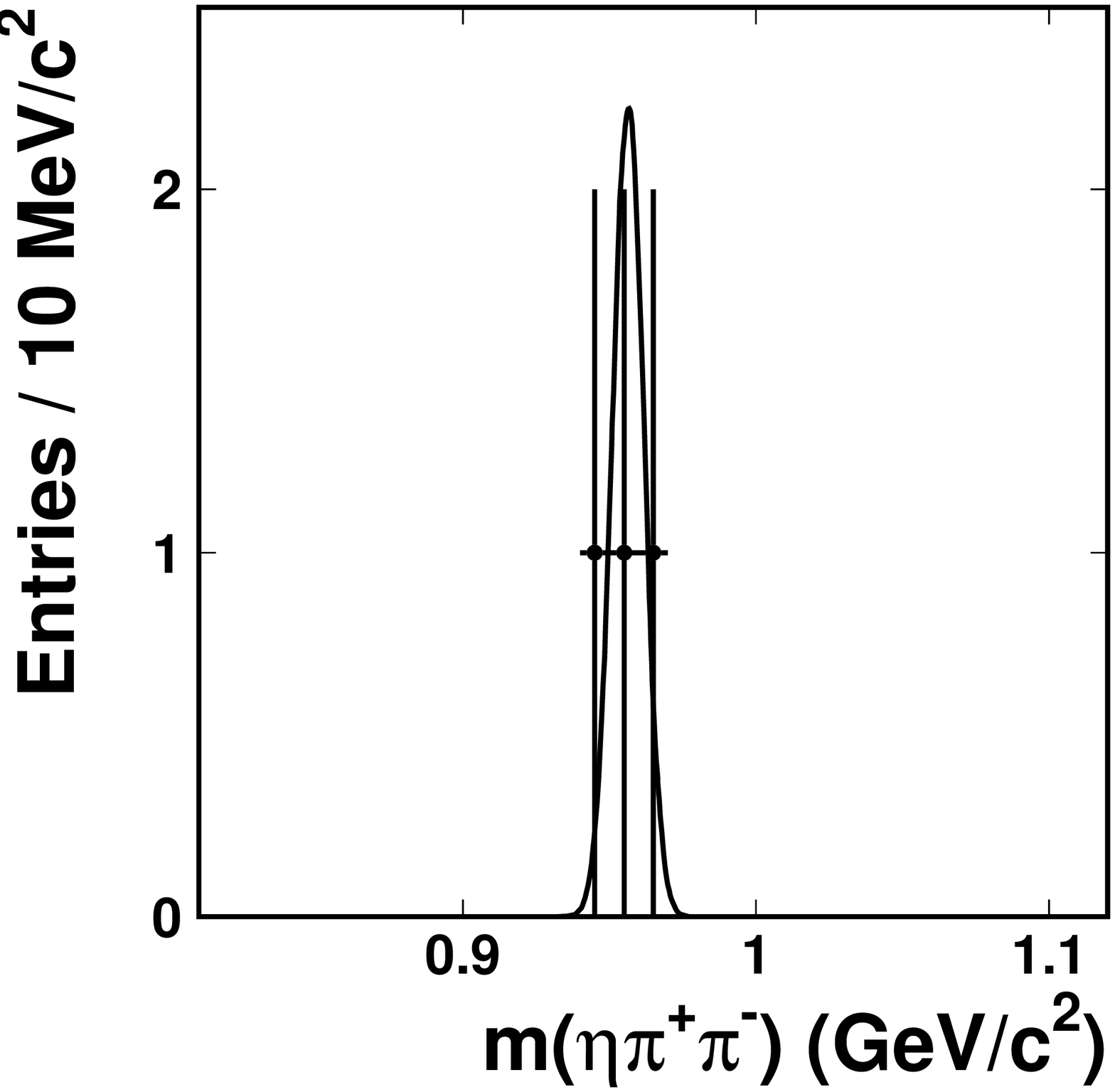}
\caption{The mass projections for the scatter plot
$m(K^+K^-)$ vs. $m(\eta\pi^+\pi^-)$ onto
{\bf a)} $K^+K^-$ and
{\bf b)} $\eta\pi^+\pi^-$
for the reaction $e^+e^- \rightarrow \phi\eta' \rightarrow K^+K^-\eta\pi^+\pi^-$.
The solid curves show the result of the two-dimensional fit,
the dotted curves show the background contamination.}
\end{figure}

\begin{figure}
\vspace*{\baselineskip}
{\hspace*{0.12\columnwidth}\bf a) \hspace*{0.42\columnwidth} b) \hfill ~} \vspace*{-2\baselineskip} \\
\includegraphics[width=0.48\columnwidth]{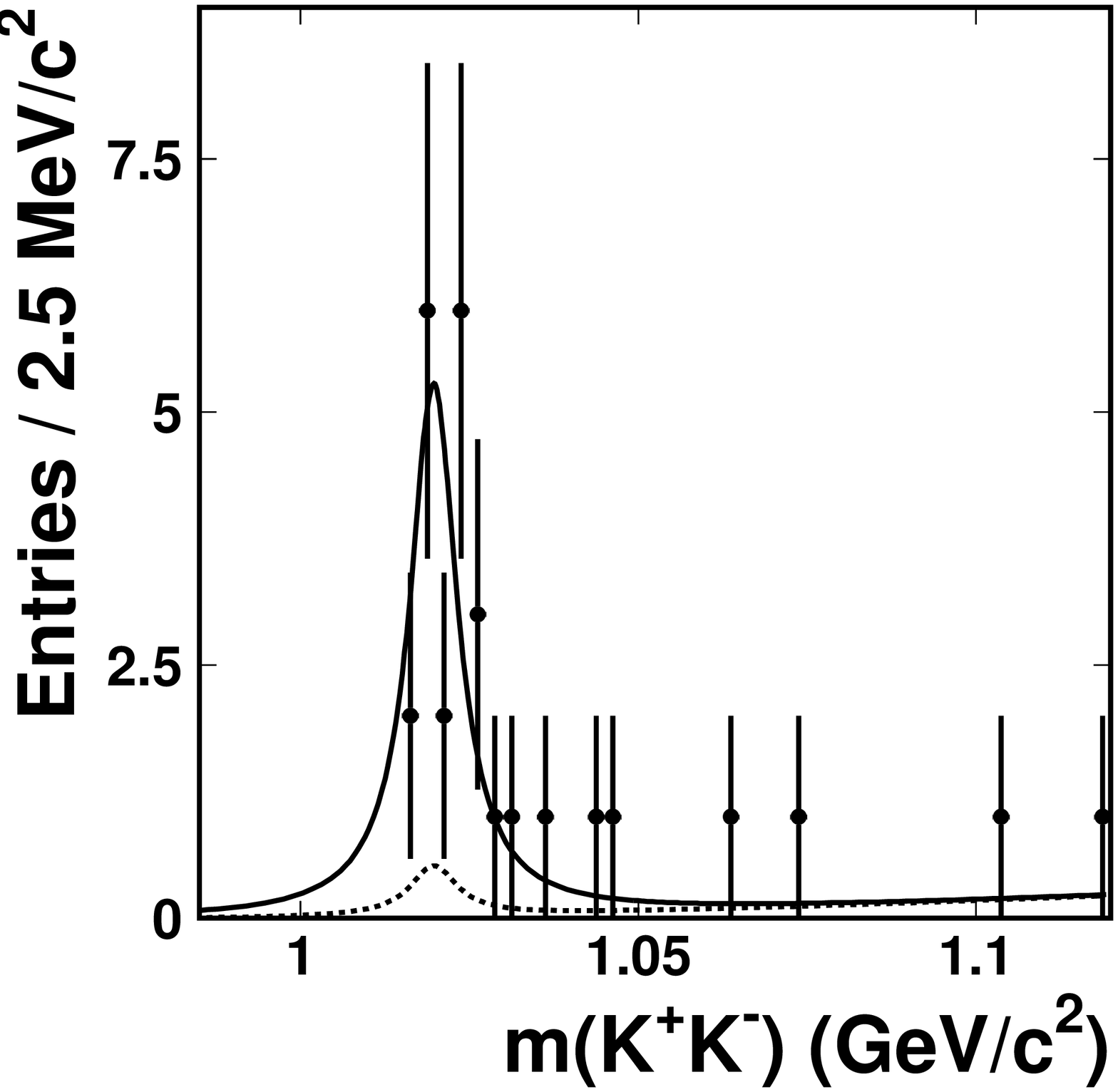}
\includegraphics[width=0.48\columnwidth]{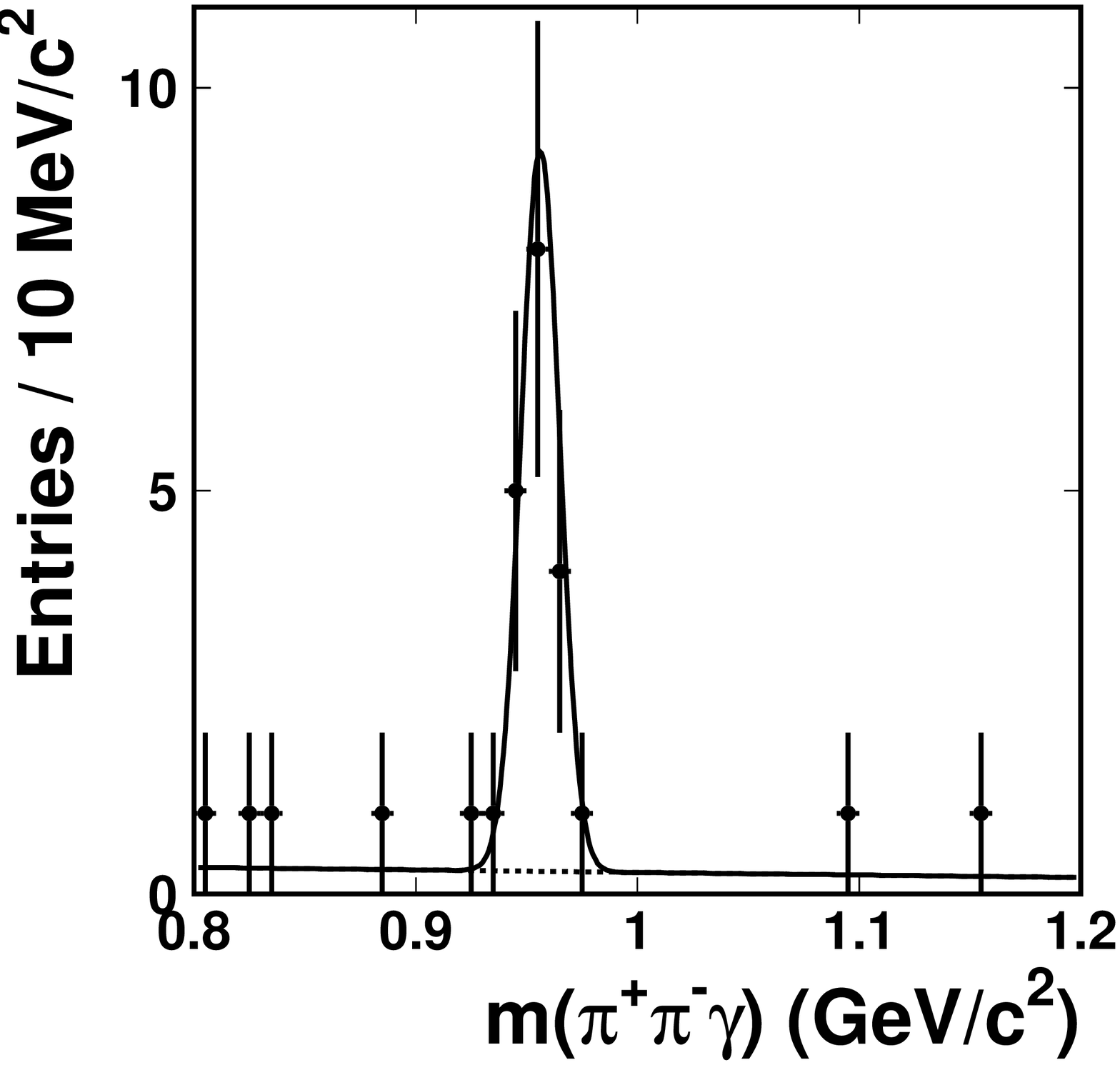}
\caption{The mass projections for the scatter plot
$m(K^+K^-)$ vs. $m(\pi^+\pi^-\gamma)$ onto
{\bf a)} $K^+K^-$ and
{\bf b)} $\pi^+\pi^-\gamma$
for the reaction $e^+e^- \rightarrow \phi\eta' \rightarrow K^+K^-\pi^+\pi^-\gamma$.
The solid curves show the result of the two-dimensional fit,
the dotted curves show the background contamination.}
\end{figure}

\begin{figure}
\vspace*{\baselineskip}
{\hspace*{0.12\columnwidth}\bf a) \hspace*{0.42\columnwidth} b) \hfill ~} \vspace*{-2\baselineskip} \\
\includegraphics[width=0.48\columnwidth]{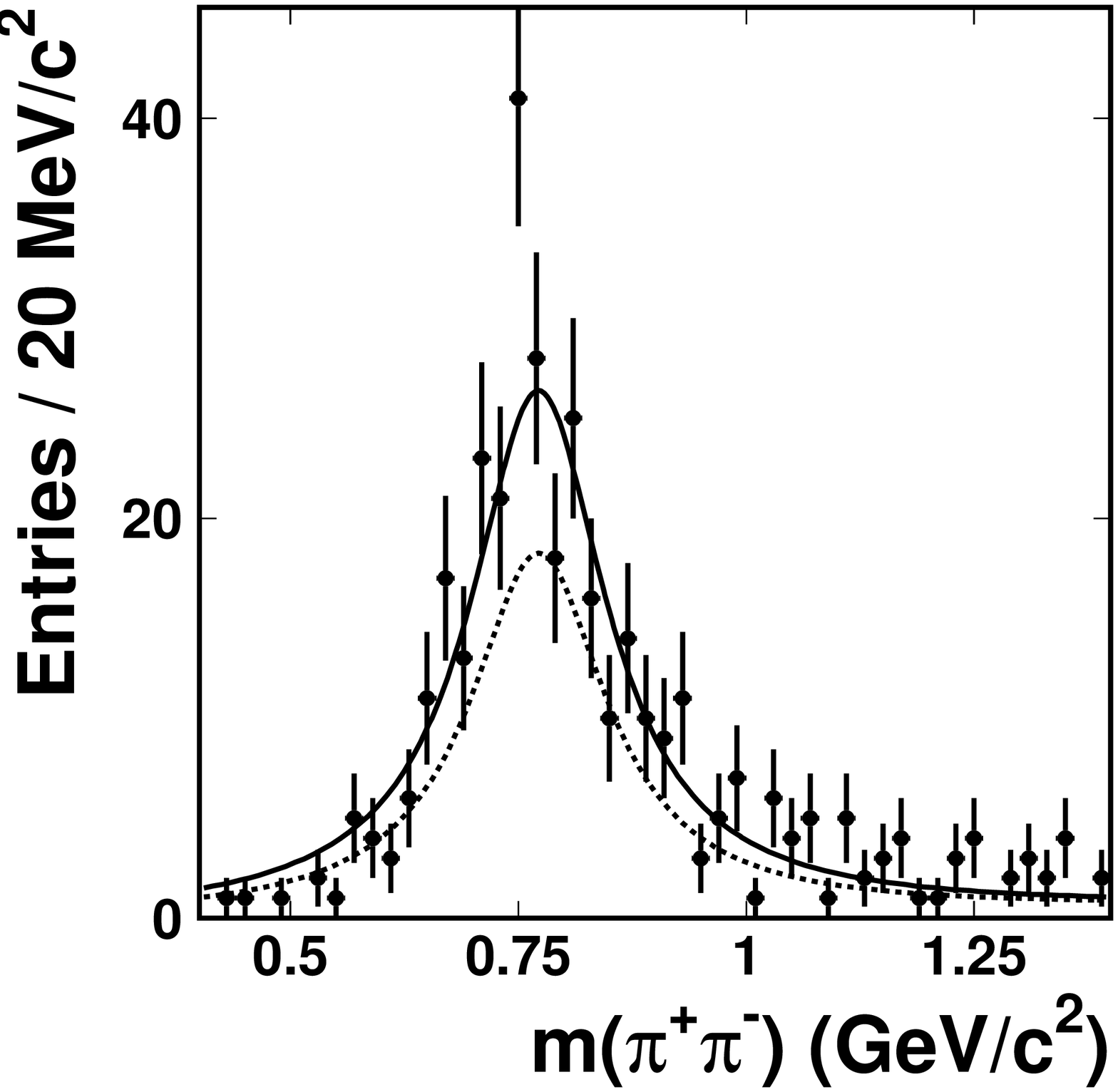}
\includegraphics[width=0.48\columnwidth]{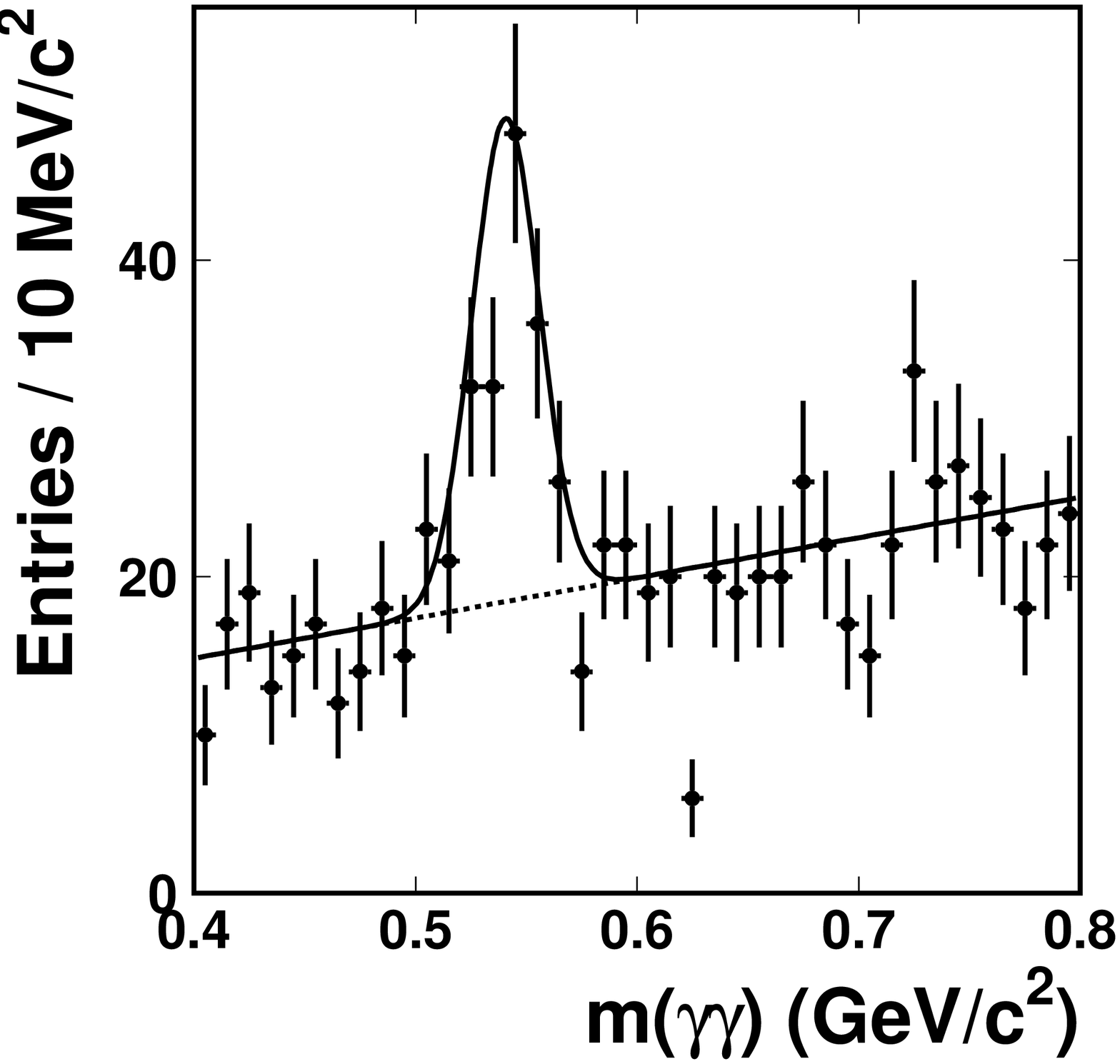}
\caption{The mass projections for the scatter plot
$m(\pi^+\pi^-)$ vs. $m(\gamma\gamma)$ onto
{\bf a)} $\pi^+\pi^-$ and
{\bf b)} $\gamma\gamma$
for the reaction $e^+e^- \rightarrow \rho\eta \rightarrow \pi^+\pi^-\gamma\gamma$.
The solid curves show the result of the two-dimensional fit,
the dotted curves show the background contamination.}
\end{figure}

\begin{figure}
\vspace*{\baselineskip}
{\hspace*{0.12\columnwidth}\bf a) \hspace*{0.42\columnwidth} b) \hfill ~} \vspace*{-2\baselineskip} \\
\includegraphics[width=0.48\columnwidth]{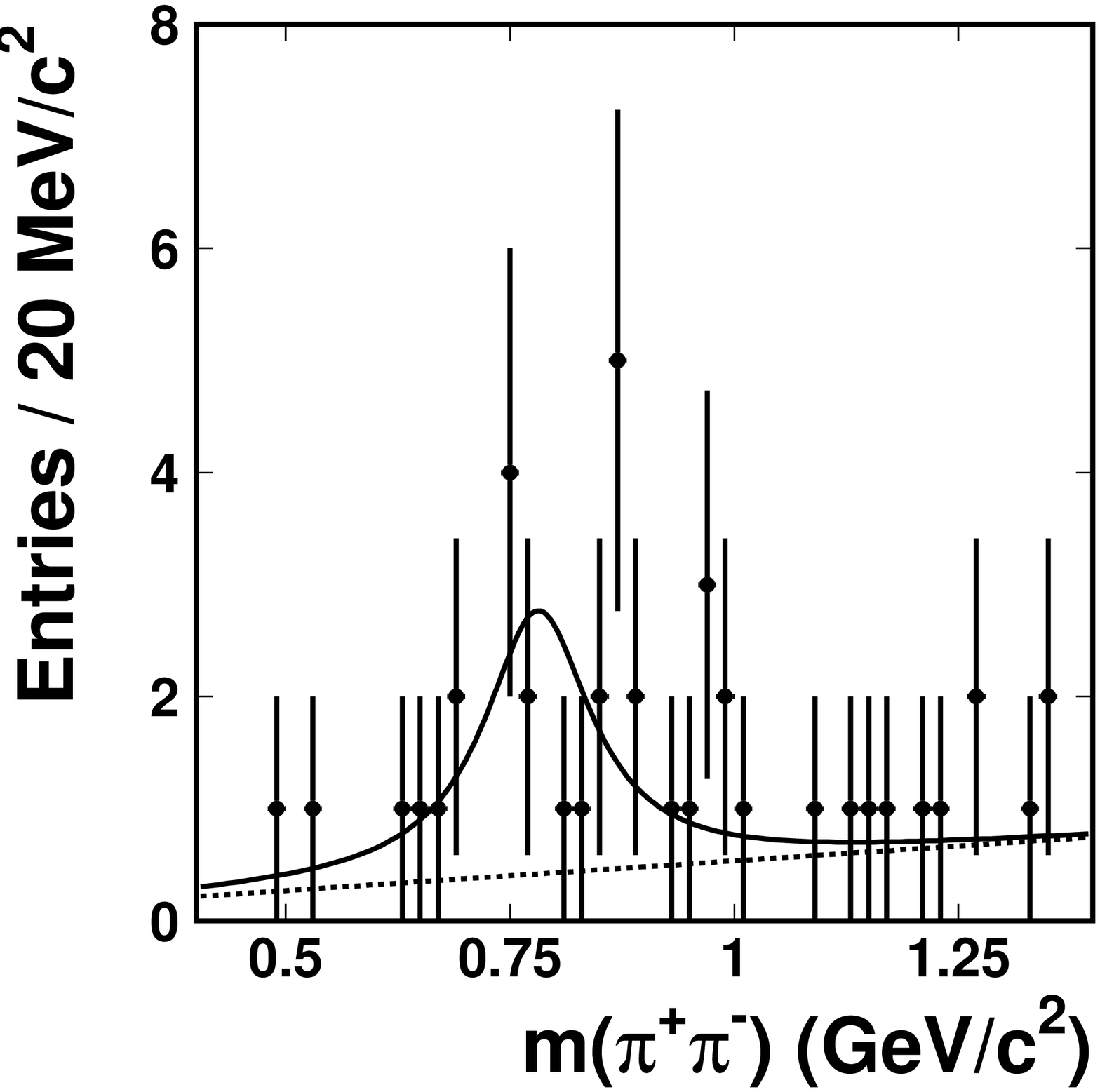}
\includegraphics[width=0.48\columnwidth]{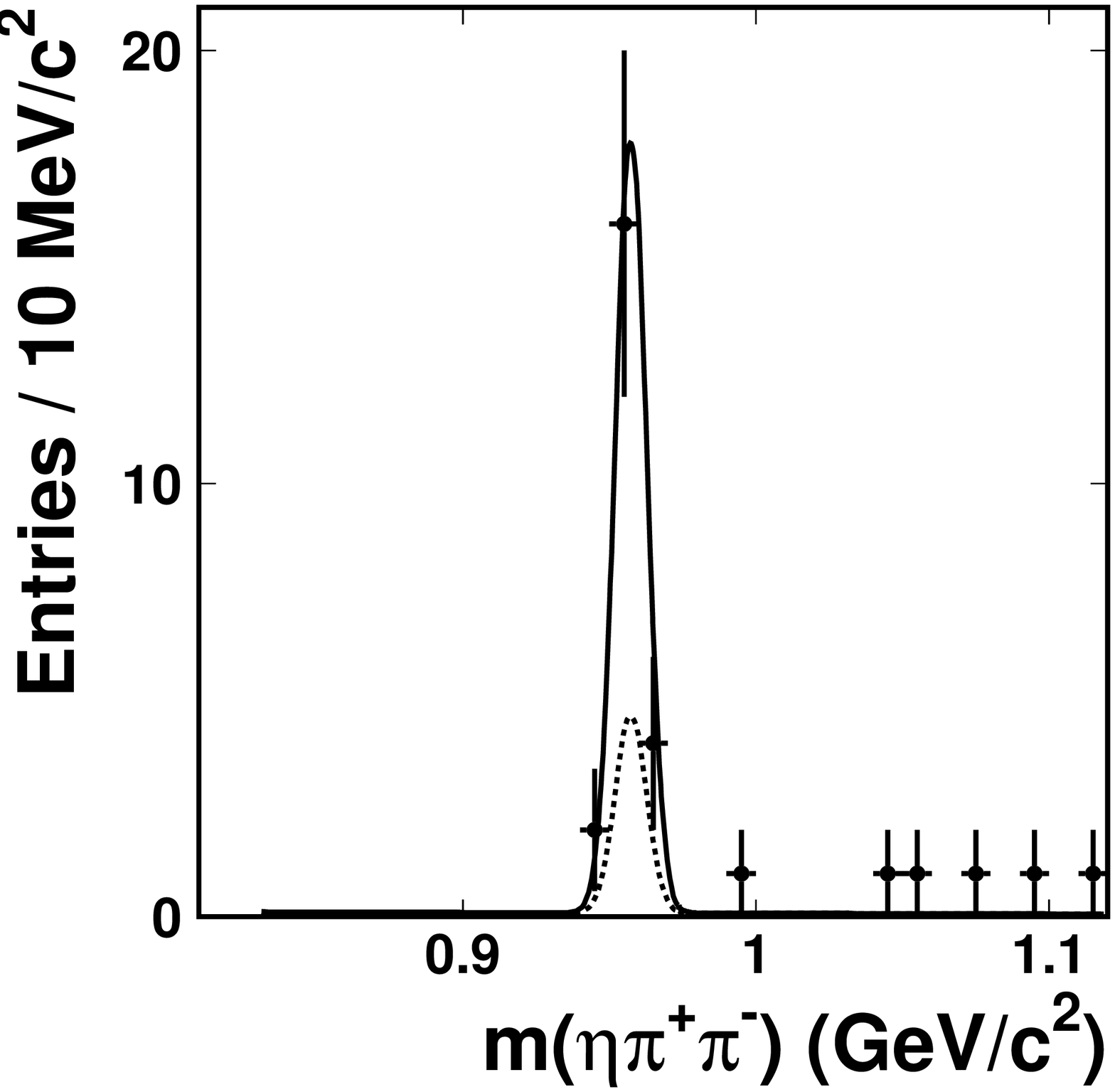}
\caption{The mass projections for the scatter plot
$m(\pi^+\pi^-)$ vs. $m(\eta\pi^+\pi^-)$ onto
{\bf a)} $\pi^+\pi^-$ and
{\bf b)} $\eta\pi^+\pi^-$
for the reaction $e^+e^- \rightarrow \rho\eta' \rightarrow \pi^+\pi^-\eta\pi^+\pi^-$.
The solid curves show the result of the two-dimensional fit,
the dotted curves show the background contamination.}
\end{figure}

\begin{figure}
\vspace*{\baselineskip}
{\hspace*{0.12\columnwidth}\bf a) \hspace*{0.42\columnwidth} b) \hfill ~} \vspace*{-2\baselineskip} \\
\includegraphics[width=0.48\columnwidth]{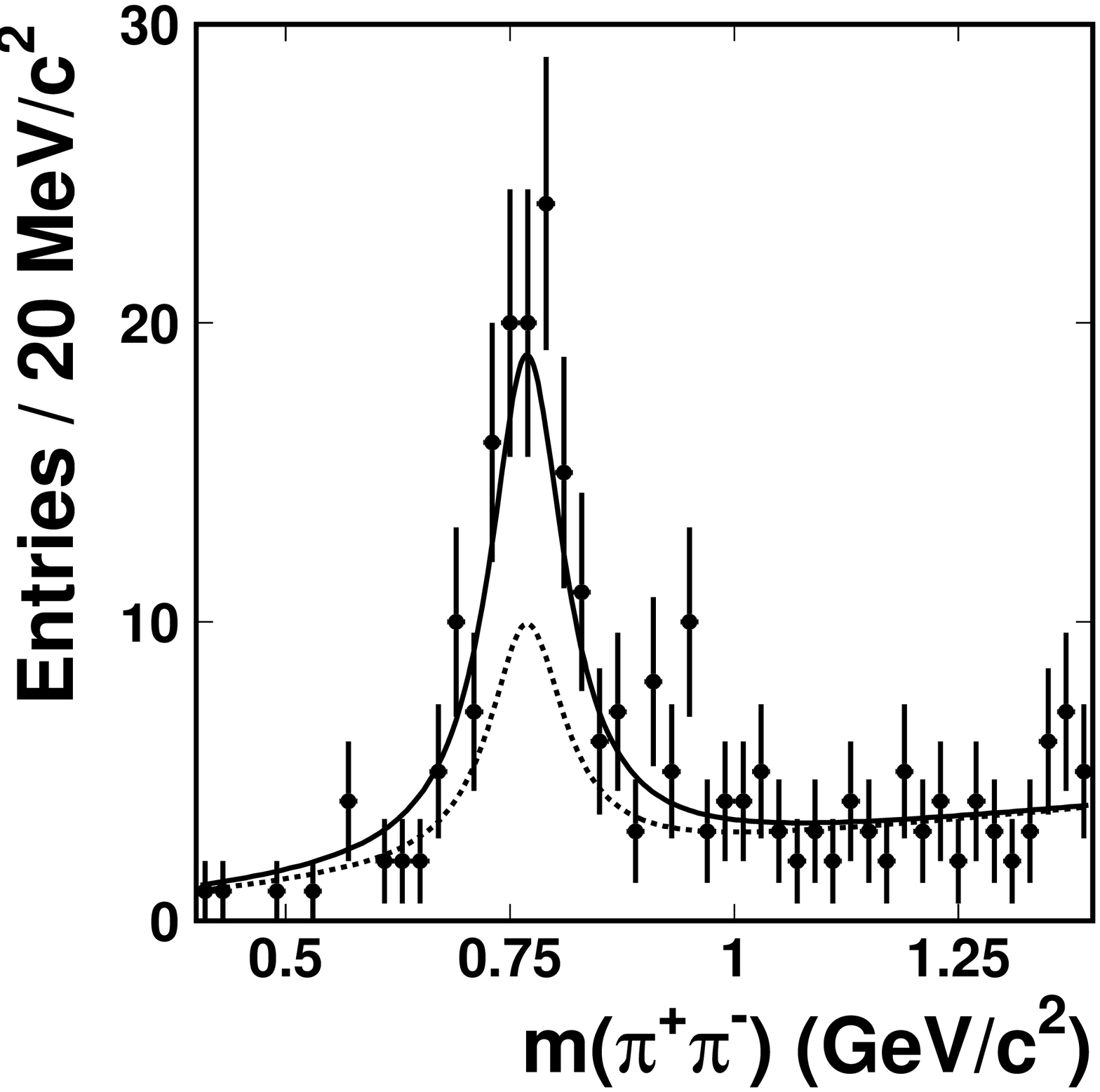}
\includegraphics[width=0.48\columnwidth]{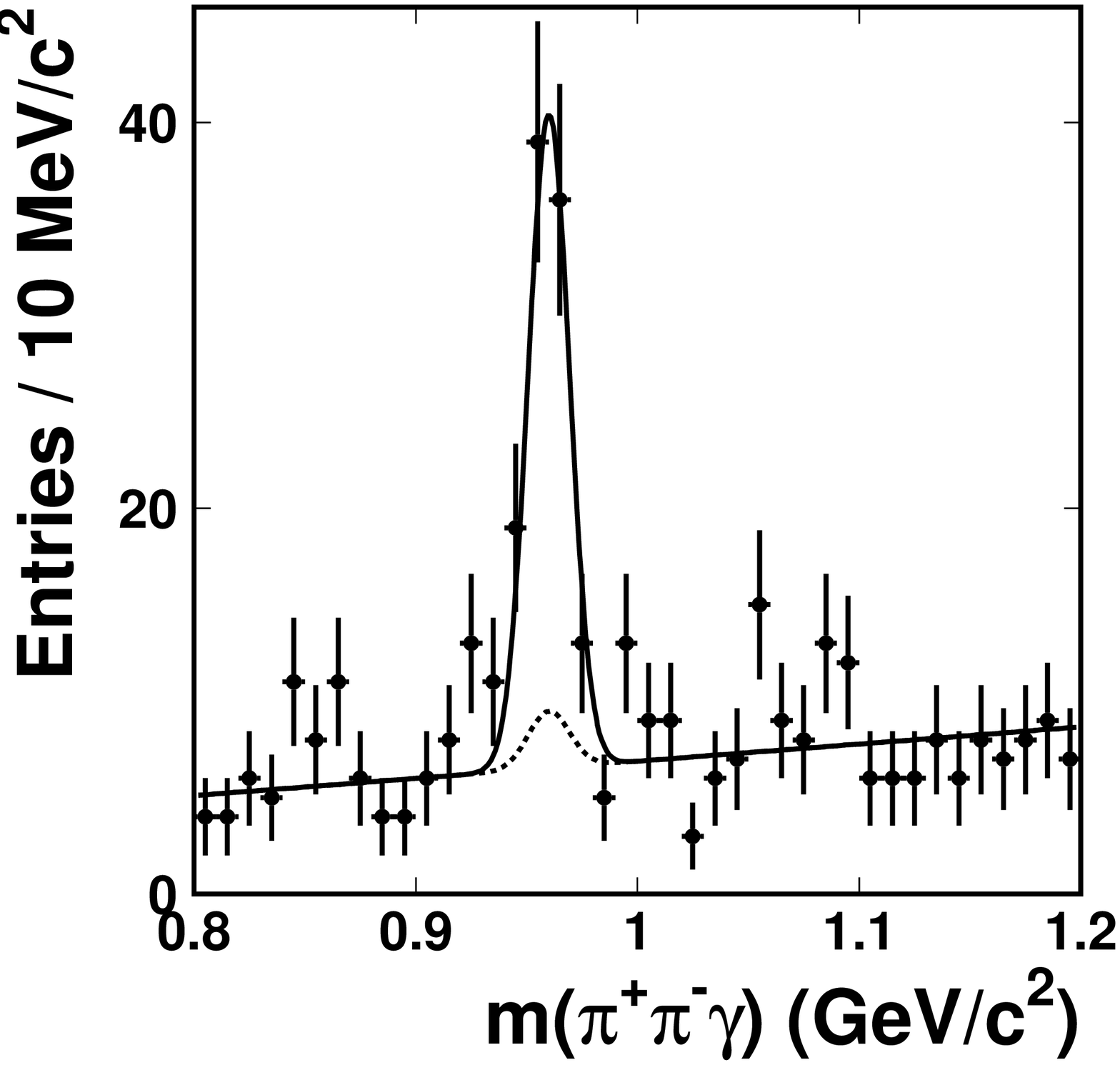}
\caption{The mass projections for the scatter plot
$m(\pi^+\pi^-)$ vs. $m(\pi^+\pi^-\gamma)$ onto
{\bf a)} $\pi^+\pi^-$ and
{\bf b)} $\pi^+\pi^-\gamma$
for the reaction $e^+e^- \rightarrow \rho\eta' \rightarrow \pi^+\pi^-\pi^+\pi^-\gamma$.
The solid curves show the result of the two-dimensional fit,
the dotted curves show the background contamination.}
\end{figure}

The corresponding one-dimensional projections together with the results of a
two-dimensional fit for different reactions are shown in Figs.~2(a,b)-7(a,b).
From the fitted signal yields, $N$, we determine the corresponding cross
sections according to the formula:
\begin{equation}
\sigma = \frac{N}{L\, B_V\, B_P\, \varepsilon},
\end{equation}
where $L$ is the integrated luminosity (516 fb$^{-1}$), $B_V$, $B_P$ are the
branching fractions of the corresponding decay channels of the vector and
pseudoscalar mesons~\cite{PDG} and $\varepsilon$ is the corresponding detection
efficiency. The detection efficiencies are determined from Monte Carlo samples
where the $e^+e^- \rightarrow V P$ reactions are generated without ISR and with
an angular dependence corresponding to a $J^P=1^-$ initial state~\cite{Chung}:
\begin{eqnarray}
\lefteqn{\frac{dN}{d\cos\theta^*d\cos\theta_Vd\phi_V}\propto}\nonumber\\
&&\sin^2\theta_V(1+\cos^2\theta^*+\cos2\phi_V\sin^2\theta^*),
\end{eqnarray}
with the production angle $\theta^*$ defined as the angle between the vector
meson direction and incident $e^-$ beam in the CM frame. The vector meson
helicity angle  $\theta_V$ is defined as the polar angle, measured in the
vector meson rest frame, of the positive decay product momentum direction with
respect to an axis that is aligned with the vector meson momentum direction in
the CM frame. The variable $\phi_V$ is the vector meson's positive decay
product azimuthal angle around the direction of the vector meson measured with
respect to the plane formed by the vector meson and the incoming electron. The
generated events were passed through a full GEANT~\cite{GEANT} Belle simulation
and reconstruction procedures including the trigger simulation. The trigger
efficiencies, estimated with MC samples described above, are about 93\% and
over 97\% for two-charged-track and four-charged-track events, respectively.

The cross sections before applying radiative corrections together with the
observed numbers of signal events,  significances of the fit and efficiencies
are presented in Table~\ref{tab:result}. Efficiencies for the processes
$e^+e^- \rightarrow \phi \eta' (\eta \pi^+\pi^-)$ and 
$e^+e^- \rightarrow \rho \eta' (\eta \pi^+\pi^-)$ include the branching
fraction of
$\eta \rightarrow \gamma \gamma \quad (39.31\pm 0.20)\%$~\cite{PDG}.

\begin{table}
\caption{Observed number of events, significance ($\Sigma$) of the fit,
efficiencies and cross sections ($\sigma$).
\label{tab:result}}
\begin{center}{\small \begin{tabular}{l|r|c|r|r} \hline \hline
{\bf Process} & {\hfill\bf $N_\mathrm{signal}$\hfill ~} & {\bf $\Sigma$} & {\hfill\bf $\varepsilon$, \%\hfill ~} & {\hfill\bf $\sigma$, fb\hfill ~}\\ \hline
$\phi \eta(\gamma\gamma)$      & $ 14.6\pm  4.3$ &  8.0 & 14.1  & $1.1 \pm 0.3$\\
$\phi \eta'(\eta\pi^+\pi^-)$   & $  3.0\pm  1.7$ & 12.0 & 0.917 & $2.9 \pm 1.6$\\
$\phi \eta'(\pi^+\pi^-\gamma)$ & $ 19.6\pm  4.5$ & 30.0& 5.36  & $4.9 \pm 1.1$\\
$\phi \eta'$(comb.)            &                 &&       & $4.3 \pm 0.9$\\  
$\rho \eta(\gamma\gamma)$      & $116.3\pm 20.2$ &  9.2 & 23.2  & $2.5 \pm 0.4$\\
$\rho \eta'(\eta\pi^+\pi^-)$   & $ 17.9\pm  4.8$ &  7.9 & 3.58  & $2.2 \pm 0.6$\\
$\rho \eta'(\pi^+\pi^-\gamma)$ & $ 72.1\pm 15.0$ &  7.6 & 14.3  & $3.3 \pm 0.7$\\ 
$\rho \eta'$(comb.)            &                 &      &       & $2.7 \pm 0.5$\\ \hline \hline
\end{tabular}}\end{center}
\end{table}

The systematic uncertainty on the $\eta \rightarrow \gamma\gamma$ detection
efficiency is dominated by the uncertainty on the MC shower simulation in the
ECL and other material. In order to estimate this uncertainty, we compare the
ratio of signal yields for the decays $\eta \rightarrow \gamma\gamma$ and
$\eta \rightarrow \pi^0\pi^0\pi^0$ in the data and in Monte Carlo simulation.
We observe a difference of about 4\%, which is treated as the systematic error
in the $\eta\rightarrow\gamma\gamma$ detection efficiency. We assume the
uncertainty in the single photon detection efficiency to be 2\%. The
systematic uncertainties due to the experimenta errors in the branching
fractions of the analyzed decay channels~\cite{PDG} are 1.3\%, 3.4\%, 0.5\%
and 3.2\% for the $\phi\eta$, $\phi\eta'$, $\rho\eta$ and $\rho\eta'$,
respectively. The systematic uncertainty in the tracking efficiency is
estimated from $D^{*+} \rightarrow D^0 \pi^+ \rightarrow K^-\pi^+\pi^+$ decays
to be 1\% per track. The systematic uncertainty from the two-dimensional fit is
estimated from variations in the number of events with mass values and widths
floating and fixed. It is estimated to be 1.5\%. The uncertainty in the
luminosity measurement is determined by the accuracy of the Bhabha generator,
which is 1.4\%. The systematic uncertainty due to trigger efficiency is
obtained from comparison of the rate of the $e^+e^- \rightarrow \phi\gamma$
events with the one expected from MC simulation. It is taken to be 1\%.
The uncertainty due to limited Monte Carlo statistics is at most 2\% for the
process $e^+e^- \rightarrow \phi\eta' \rightarrow K^+K^-\gamma\gamma\pi^+\pi^-$.
The uncertainty on the charged kaon identification is estimated by comparing
efficiencies of kaon identification in decays
$D^{*+} \rightarrow D^0 \pi^+ \rightarrow K^-\pi^+\pi^+$ for the data and Monte
Carlo events. For the uncertainty of a charged kaon identification we take the
relative difference in these efficiencies, which is $0.5\%$ per kaon. The
systematic uncertainties for all analyzed channels are given in
Table~\ref{tab:error}.

\begin{table}
\caption{Total systematic uncertainties for analyzed channels.
\label{tab:error}}
\begin{center}\begin{tabular}{l|c} \hline \hline
{\bf Channel} & {\bf Error (\%)}\\ \hline
$e^+e^-\rightarrow\phi\eta \rightarrow K^+K^-\gamma\gamma$               & 5.3 \\
$e^+e^-\rightarrow\phi\eta'\rightarrow K^+K^-\gamma\gamma\pi^+\pi^-$     & 7.4 \\
$e^+e^-\rightarrow\phi\eta'\rightarrow K^+K^-\pi^+\pi^-\gamma$           & 6.2 \\
$e^+e^-\rightarrow\rho\eta \rightarrow \pi^+\pi^-\gamma\gamma$           & 5.0 \\
$e^+e^-\rightarrow\rho\eta'\rightarrow \pi^+\pi^-\gamma\gamma\pi^+\pi^-$ & 7.0 \\
$e^+e^-\rightarrow\rho\eta'\rightarrow \pi^+\pi^-\pi^+\pi^-\gamma$       & 5.9 \\ \hline \hline
\end{tabular}\end{center}
\end{table}

To check whether the observed signals are due to $\Upsilon(4S)$ decays, we
scale the off-resonance signals to the on-resonance luminosity, and subtract
them from the on-resonance signals. The observed numbers of events in the
off-resonance data are $1 \pm 1$, $1 \pm 1$, $2 \pm 1.4$, $15.2 \pm 4.7$,
$1 \pm 1$, $7 \pm 3.6$ for the processes $\phi \eta(\gamma\gamma)$,
$\phi \eta'(\eta\pi^+\pi^-)$, $\phi \eta'(\pi^+\pi^-\gamma)$,
$\rho \eta(\gamma\gamma)$, $\rho \eta'(\eta\pi^+\pi^-)$ and
$\rho \eta'(\pi^+\pi^-\gamma)$, respectively. The resulting branching fractions
for  $\Upsilon(4S) \rightarrow VP$ are $(0.4\pm0.8)\times 10^{-6}$,
$(-0.6\pm2.8)\times 10^{-6}$, $(-0.5\pm1.0)\times 10^{-6}$,
$(0.8\pm0.9)\times 10^{-6}$ for the $\phi\eta$, $\phi\eta'$, $\rho\eta$,
$\rho\eta'$ channels, respectively, which are consistent with zero. These
results can be expressed as the 90\% confidence level upper limits~\cite{FC},
which are equal to $1.8 \times 10^{-6}$, $4.3 \times 10^{-6}$,
$1.3 \times 10^{-6}$, $2.5 \times 10^{-6}$ for the $\phi\eta$, $\phi\eta'$,
$\rho\eta$, $\rho\eta'$ channels, respectively. The systematic uncertainties
are also taken into account for upper limit calculations

%
In the light cone approach the authors of Refs.~\cite{LC, Likhoded} gave
predictions for the cross sections of the reactions analyzed by Belle at
$\sqrt{s}=10.58\GeV$. In Table~\ref{tab:theory} we present the Belle cross
sections radiatively corrected according to Ref.~\cite{Eidelman} together with
theory predictions~\cite{LC, Likhoded}. The radiatively corrected cross section
can be written as
\begin{equation}
\sigma_0=\frac{\sigma}{1+\delta} ,
\end{equation}
where $\sigma$ is taken from equation (1). The value of $1+\delta$
corresponding to the energy cut of 0.3 GeV is equal to 0.809~\cite{Eidelman}.
The BaBar measurement of the reaction
$e^+e^- \rightarrow \phi\eta$~\cite{BaBar1} is also presented in the table.
The corresponding value of $1+\delta$ for the BaBar energy cut of 0.23 GeV is
0.768. BaBar reports that the cross section of the reaction
$e^+e^- \rightarrow \phi\eta$ is $2.9 \pm 0.5 \pm 0.1$~fb~\cite{BaBar1}. The
Belle cross section is smaller than the BaBar result by about $2.3\sigma$.

\begin{table*}
\caption{The values of cross sections of reactions $e^+e^- \rightarrow VP$,
radiatively corrected according to Ref.~\cite{Eidelman}, measured by
Belle and predicted by theory~\cite{LC, Likhoded} and the BaBar
measurement.
\label{tab:theory}}
\begin{center}\begin{tabular}{l|r|r|c|c} \hline \hline
{\bf Channel} &{\bf $\sigma_0$ Belle (fb)} & {\bf $\sigma$~\cite{LC} (fb)} &{\bf $\sigma$~\cite{Likhoded} (fb)} & {\bf $\sigma_0$ BaBar (fb)}\\ \hline
$e^+e^- \rightarrow \phi \eta$  & $1.4 \pm 0.4 \pm 0.1$ & $3.3-4.3$ & $2.4-3.4$ & $2.9 \pm 0.5 \pm 0.1$\\
$e^+e^- \rightarrow \phi \eta'$ & $5.3 \pm 1.1 \pm 0.4$ & $4.4-5.8$ & $3.5-5.0$ & -- \\
$e^+e^- \rightarrow \rho \eta$  & $3.1 \pm 0.5 \pm 0.1$ & $2.4-3.1$ & $2.4-3.5$ & -- \\
$e^+e^- \rightarrow \rho \eta'$ & $3.3 \pm 0.6 \pm 0.2$ & $1.5-2.1$ & $1.6-2.3$ & -- \\ \hline \hline
\end{tabular}\end{center}
\end{table*}

From Table~\ref{tab:theory} we see that in comparison to
theory~\cite{LC, Likhoded} the Belle experimental cross section for
$e^+e^- \rightarrow \phi \eta$ is significantly lower,
$e^+e^- \rightarrow \rho\eta'$ is about $1.8\sigma$ higher, while
$e^+e^- \rightarrow \phi\eta'$ and $e^+e^- \rightarrow \rho\eta$ agree within
errors with theory. There is also a discrepancy between the data and light cone
expectation in the ratio of the cross section of $\eta$ meson production
together with vector mesons to that of $\eta'$ production. As can be seen from
Table~\ref{tab:theory}, the light cone approach~\cite{LC} predicts
$\frac{\sigma(e^+e^-\rightarrow\rho\eta)}{\sigma(e^+e^-\rightarrow\rho\eta')}
> 1$ while this trend is not observed in data.

\begin{figure}
\includegraphics[width=0.9\columnwidth]{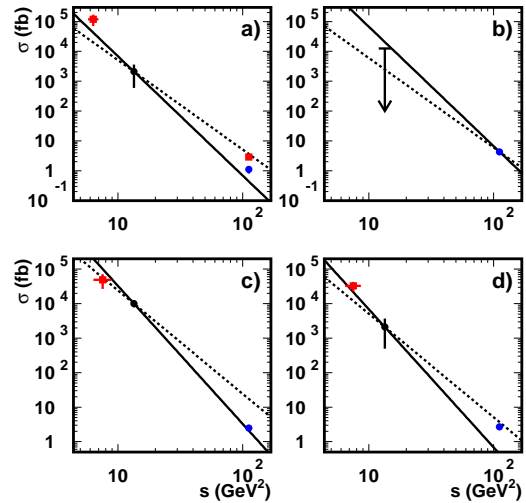}
\caption{\label{fig:sdep}
The measured cross sections at $\sqrt{s}\approx2.5,\,2.75\GeV$ by BaBar,
$\sqrt{s} = 3.67 $ GeV by CLEO and at $\sqrt{s} = 10.58$ GeV by BaBar and
Belle for the various processes. BaBar measurements are represented by squares.
{\bf a)} $e^+e^- \rightarrow \phi \eta$,
{\bf b)} $e^+e^- \rightarrow \phi \eta'$;
{\bf c)} $e^+e^- \rightarrow \rho \eta$;
{\bf d)} $e^+e^- \rightarrow \rho \eta'$.
In the plot b) the CLEO upper limit is shown by the arrow. The solid lines
correspond to a $1/s^4$ dependence and the dashed ones represent $1/s^3$.}
\end{figure}

The energy dependence of the cross sections may have important theoretical
implications. In Figs.~\ref{fig:sdep}(a-d) we show the Belle data radiatively
corrected according to Ref.~\cite{Eidelman} together with CLEO and BaBar ISR
results. The BaBar data were averaged for $\sqrt{s}$ values from $2.5$ to
$3\GeV$. We also show $1/s^3$ and $1/s^4$ dependences, which pass through the
CLEO points. In Fig.~\ref{fig:sdep}(b) the arrow shows the CLEO upper limit and
the curves pass through the Belle measurement. From Fig.~\ref{fig:sdep} we
cannot draw any definite conclusion about the energy dependence of the
$e^+e^- \rightarrow VP$ reactions.

%
To summarize, we have measured cross sections of the exclusive processes
$e^+e^- \rightarrow \phi \eta$, $e^+e^- \rightarrow \phi \eta'$,
$e^+e^- \rightarrow \rho \eta$, $e^+e^- \rightarrow \rho \eta'$ where the final
state includes soft photons with energies below $0.3\GeV$ using a 516 fb$^{-1}$
data sample recorded near $\sqrt{s} = 10.58\GeV$. The corresponding values of
the cross sections in femtobarns are: $1.4 \pm 0.4 \pm 0.1$,
$5.3 \pm 1.1 \pm 0.4$, $3.1 \pm 0.5 \pm 0.1$ and $3.3 \pm 0.6 \pm 0.2$. The
results are compared with theoretical predictions from the light cone
approach~\cite{LC, Likhoded}. Analysis of the energy dependence of the cross
sections using our results together with those of CLEO at
$\sqrt{s} = 3.67\GeV$~\cite{CLEO1} and BaBar $2.5\GeV$ for $\phi\eta$ and
$2.75\GeV$ for $\rho\eta,\quad\rho\eta'$~\cite{BaBar0} shows that there is no
universal energy dependence for these processes. The ratios of cross sections
of $\eta$ meson production together with vector mesons to the corresponding
cross sections for $\eta'$ meson production are different from the light cone
expectation~\cite{LC, Likhoded}. The 90\% confidence level upper limits on the
branching fractions of the $\Upsilon(4S) \rightarrow VP$ are
$1.8 \times 10^{-6}$, $4.3 \times 10^{-6}$, $1.3 \times 10^{-6}$,
$2.5 \times 10^{-6}$ for the $\phi\eta$, $\phi\eta'$, $\rho\eta$, $\rho\eta'$
channels, respectively.

\bigskip
We thank the KEKB group for the excellent operation of the
accelerator, the KEK cryogenics group for the efficient
operation of the solenoid, and the KEK computer group and
the National Institute of Informatics for valuable computing
and SINET3 network support.  We acknowledge support from
the Ministry of Education, Culture, Sports, Science, and
Technology (MEXT) of Japan, the Japan Society for the 
Promotion of Science (JSPS), and the Tau-Lepton Physics 
Research Center of Nagoya University; 
the Australian Research Council and the Australian 
Department of Industry, Innovation, Science and Research;
the National Natural Science Foundation of China under
contract No.~10575109, 10775142, 10875115 and 10825524; 
the Department of Science and Technology of India; 
the BK21 program of the Ministry of Education of Korea, 
the CHEP src program and Basic Research program (grant 
No. R01-2008-000-10477-0) of the 
Korea Science and Engineering Foundation;
the Polish Ministry of Science and Higher Education;
the Ministry of Education and Science of the Russian
Federation and the Russian Federal Agency for Atomic Energy;
the Slovenian Research Agency;  the Swiss
National Science Foundation; the National Science Council
and the Ministry of Education of Taiwan; and the U.S.\
Department of Energy.
This work is supported by a Grant-in-Aid from MEXT for 
Science Research in a Priority Area ("New Development of 
Flavor Physics"), and from JSPS for Creative Scientific 
Research ("Evolution of Tau-lepton Physics").

\end{document}

%% file: author.tex
  \author[Protvino]{K.~Belous} 
  \author[Protvino]{M.~Shapkin} 
  \author[KEK]{I.~Adachi} 
  \author[Tokyo]{H.~Aihara} 
  \author[BINP,Novosibirsk]{K.~Arinstein} 
  \author[BINP,Novosibirsk]{V.~Aulchenko} 
  \author[Sydney]{A.~M.~Bakich} 
  \author[ITEP]{V.~Balagura} 
  \author[Melbourne]{E.~Barberio} 
  \author[KEK]{W.~Bartel} 
% ...

\address[BINP]{Budker Institute of Nuclear Physics, Novosibirsk, Russian Federation}
% ...
\address[KEK]{High Energy Accelerator Research Organization (KEK), Tsukuba, Japan}
% ...
\address[Protvino]{Institute for High Energy Physics, Protvino, Russian Federation}
% ...
\address[ITEP]{Institute for Theoretical and Experimental Physics, Moscow, Russian Federation}
% ...
\address[Melbourne]{University of Melbourne, Victoria, Australia}
% ...
\address[Novosibirsk]{Novosibirsk State University, Novosibirsk, Russian Federation}
% ...
\address[Sydney]{University of Sydney, Sydney, NSW, Australia}
% ...
\address[Tokyo]{Department of Physics, University of Tokyo, Tokyo, Japan}
% ...